\documentclass[12pt]{article}
\usepackage{amsmath,amssymb,amsthm,amsfonts,url}

\newcommand{\I}{\makebox[0pt][l]{1}\hspace{0.3ex}\mbox{1}}
\newcommand{\ket}[1]{\vert#1\rangle}

\newcommand{\braket}[2]{\langle #1 \vert #2\rangle}
\newcommand{\sandwich}[3]{\mbox{$ \langle #1 | #2 | #3 \rangle $}}

\renewcommand{\mod}{\text{mod}}

\newcommand{\x}{^{(x)}}
\newcommand{\y}{^{(y)}}
\newcommand{\0}{^{(0)}}
\newcommand{\1}{^{(1)}}
\newcommand{\2}{^{(2)}}
\newcommand{\ba}{\begin{eqnarray}}
\newcommand{\ea}{\end{eqnarray}}

\newtheorem{theo}{Theorem}
\newtheorem{defi}{Definition}
\newtheorem{prop}[theo]{Proposition}

\newtheorem{corol}[theo]{Corollary}

\begin{document}
\title{Pseudo-telepathy: input cardinality and Bell-type inequalities}
\author{
Nicolas Gisin \quad Andr\'e Allan M\'ethot \quad Valerio Scarani\\[0.5cm]
\normalsize\sl Group of Applied Physics\\[-0.1cm]
\normalsize\sl Universit\'e de Gen\`eve, Rue de l'\'Ecole-de-M\'edecine 20\\[-0.1cm]
\normalsize\sl CH-1211 Gen\`eve 4~~\textsc{Switzerland}\\
\normalsize\url{{nicolas.gisin,andre.methot,valerio.scarani}@physics.unige.ch}}
\date{11 October 2006}
\maketitle

\begin{abstract}
Pseudo-telepathy is the most recent form of rejection of locality.
Many of its properties have already been discovered: for instance,
the minimal entanglement, as well as the minimal cardinality of
the output sets, have been characterized. This paper contains two
main results. First, we prove that no bipartite pseudo-telepathy
game exists, in which one of the partners receives only two
questions; as a corollary, we show that the minimal ``input
cardinality'', that is, the minimal number of questions required
in a bipartite pseudo-telepathy game, is $3\times 3$. Second, we
study the Bell-type inequality derived from the pseudo-telepathy
game known as the Magic Square game: we demonstrate that it is a
tight inequality for 3 inputs and 4 outputs on each side and
discuss its weak resistance to noise.
\end{abstract}


\section{Introduction}\label{sec:intro}
Implicitly present in the works of Heywood and Redhead \cite{hr83}
and of Greenberger, Horne, Zeilinger and Mermin
\cite{ghz89,mermin90a}, the notion of {\em pseudo-telepathy} was
first explicitly introduced by Brassard, Cleve and
Tapp~\cite{bct99}. In this work, the authors used a game which was
possible to win classically only if the players exchanged
information, while quantum players sharing entanglement could
dispense with communication altogether. They used their result to
show a lower bound on the number of bits of communication classical
players need to exchange in order to simulate quantum measurements
on many maximally entangled pairs of qubits. Since no communication
whatsoever is required in the quantum strategy to the game, we can
see that their game is actually a proof against locality in the same
sense as the result of Bell~\cite{bell64}. Furthermore, it was shown
recently that pseudo-telepathy is a stronger rejection of the
locality assumption than the traditional Bell theorems, or even the
Hardy~\cite{hardy92} theorem~\cite{bmt04,methot05}. We will give a
complete definition of pseudo-telepathy in Section~\ref{sec:defs},
for a survey of known results see~\cite{bbt04a}.

Much work has been done in order to characterize pseudo-telepathy.
A question of interest is ``What kind of correlations are required
in order to give rise to a pseudo-telepathy game?'' The minimal
entanglement necessary in order to have a pseudo-telepathy game
was resolved in~\cite{bmt04} where the authors have shown that
either a pair of entangled qutrits or three entangled qubits are
needed. In~\cite{chtw04}, the minimal cardinality of the outputs
sets was uncovered: in the bipartite scenario, at least one player
must have more than two possible outputs, while in the
multipartite scenario two outputs per player are enough. In
Section \ref{sec:min} we provide the answer to the last question
along these lines that has remained open: how many possible
questions must one incorporate in a game for it to be a
pseudo-telepathy game.

In Section \ref{sec:msg}, we consider pseudo-telepathy in the
context of the characterization of non-local correlations. In the
last years, the understanding of non-locality has been significantly
improved using a geometric view of probability distributions: local
distributions form a closed convex set with a finite number of
extremal points (a ``polytope''), whose facets define tight
Bell-type inequalities \cite{pito,cg03}. It is interesting to ask
whether pseudo-telepathy games are associated to tight Bell
inequalities: while a general proof or a counter-example are still
missing, the answer is positive for the three-partite Mermin-GHZ
game~\cite{mermin90a} as proved in~\cite{proofmermin} and for the
bipartite Magic Square game~\cite{aravind02} as we prove here. We
show also that the inequality derived from the Magic Square game is
less resistant to noise than another inequality, which is not
associated to a pseudo-telepathy game: this is not surprising, since
pseudo-telepathy is stronger a refutation of locality.

\section{Definitions and mathematical tools}\label{sec:defs}

In this Section, we present formal definitions and the
mathematical tools that are required to demonstrate the results of
Sections \ref{sec:min} and \ref{sec:msg}.

\subsection{Pseudo-telepathy}\label{sec:pt}

In this paper, we choose the following notations:\\
Alice's input: $x\in{\cal X}=\{0,1,...,m_A-1\}$;\\
Alice's output: $a\in{\cal A}=\{0,1,...,n_A-1\}$;\\
Bob's input: $y\in{\cal Y}=\{0,1,...,m_B-1\}$;\\
Bob's output: $b\in{\cal B}=\{0,1,...,n_B-1\}$.

\begin{defi}[Bipartite Game]
A \emph{bipartite game} $G=({\cal I}, {\cal O},W)$ is a set of
inputs ${\cal I}={\cal X}\times {\cal Y}$, a set of outputs ${\cal
O}={\cal A}\times {\cal B}$ and a relation $W\subseteq {\cal
I}\times {\cal O}$ that inputs and outputs should satisfy.
\end{defi}

\begin{defi}[Winning Strategy]
A \emph{winning strategy} for a bipartite game  $G=({\cal I},
{\cal O},W)$ is a strategy according to which for every $x\in
{\cal X}$ and $y\in {\cal Y}$, Alice and Bob output $a$ and $b$
respectively such that $(x,y,a,b)\in W$.
\end{defi}

\begin{defi}[Pseudo-telepathy]\label{def:pseudo-telepathy}
We say that a bipartite game $G$ exhibits \emph{pseudo-telepathy}
if bipartite measurements of an entangled quantum state can yield
a winning strategy, whereas no classical strategy that does not
involve communication is a winning strategy.
\end{defi}

The extension to the multipartite case is trivial, but we consider
only bipartite scenarios in this paper. The generalization of
Definition~\ref{def:pseudo-telepathy} can be translated into a set
of multipartite measurements on an entangled state where any local
classical model which is to attempt to only produce outputs that
are not forbidden by quantum mechanics will fail.

\subsection{Polytope and facets inequalities}\label{sec:poly}

For the purposes of Section \ref{sec:msg}, let's recall the
geometric formalism for probability distributions
\cite{pito,cg03}. As mentioned, we consider the situation in which
Alice receives the input $x\in \{0,1,...,m_A-1\}$ and has to
output $a\in \{0,1,...,n_A-1\}$; similarly, Bob receives the input
$y\in \{0,1,...,m_B-1\}$ and has to output $b\in
\{0,1,...,n_B-1\}$. Any probability distribution can be
represented by a vector of dimension $m_A\,m_B\,n_A\,n_B$ whose
entries are just the numbers $\Pr[a,b|x,y]$. In the space of
probability distributions, we focus on the set of local strategies
$\Omega_{m_A\,m_B\,n_A\,n_B}$ (written $\Omega_{mn}$ when
$m_A=m_B$ and $n_A=n_B$). This set is a polytope, that is, a
convex set with a finite number of extremal points. The extremal
points are deterministic local strategies, that is, points of the
form \ba \Pr[a,b|x,y]&=& \Pr[a|x]\,\Pr[b|y]\,=\,\delta_{a,a(x)}\,
\delta_{b,b(y)} \ea for all possible functions $x\mapsto a(x)$,
$y\mapsto b(y)$, and where $\delta_{i,j}$ is the Kronecker delta.
There are clearly $n_A^{m_A}\,n_B^{m_B}$ deterministic strategies.
Note that the local polytope $\Omega_{m_A\,m_B\,n_A\,n_B}$ is
embedded in a subspace of the probability space, because local
strategies automatically satisfy the so-called no-signalling
condition. It can be shown that the dimension of the space of
no-signalling probability distributions is
$d=m_Am_B(n_A-1)(n_B-1)+m_A(n_A-1)+m_B(n_B-1)$.

A polytope embedded in a $d$-dimensional space is completely
described by giving the list of its extremal points (vertices) or
equivalently, of the $(d-1)$-dimensional hypersurfaces that bound
it (facets). On the one hand, as we have just seen, it is very
easy to list the vertices. On the other hand, the characterization
in terms of facets is more useful, because it allows to decide
whether a probability point is inside the polytope, and is
therefore local; or outside it, and is then non-local; in other
words, each facet represents a tight Bell-type inequality. The
task of finding the facets given the list of vertices is
computationally hard. In Section \ref{sec:msg}, we show how a
particular pseudo-telepathy game provides a natural inequality
that is in fact a facet for the corresponding local polytope.


\section{Input Cardinality}\label{sec:min}

In this Section, we want to prove the following

\begin{theo}\label{thm:2xn}
In pseudo-telepathy, the set of possible question ${\cal X}\times
{\cal Y}$ cannot be of cardinality $2\times n$.
\end{theo}

\begin{proof}
For definiteness, we suppose that it is Alice who receives only
two questions, i.e. ${\cal X}=\{0,1\}$. We prove the statement by
{\em reductio ad absurdum}. Consider deterministic strategies: the
values $a\0$ and $a\1$ that Alice is going to output upon
receiving $x=0$ or $x=1$ are fixed. Suppose now that for a given
pair $(a\0,a\1)$ it so happens, that for all $y\in{\cal Y}$ there
exist a $b\y$ such that $(0,y,a\0,b\y)\in W$ and $(1,y,a\1,b\y)\in
W$: then there is clearly a classical winning strategy, and the
game cannot be a pseudo-telepathy game. The rest of the proof
consist in proving that exactly this situation is generated by the
most general bipartite quantum strategy with two inputs on Alice
side.

In a quantum strategy, Alice and Bob share a quantum state and
obtain the outcomes by performing measurements on the state. Since
we are not going to fix the dimension of the Hilbert space of
Alice and Bob, we can assume without loss of generality that the
measurements are projective --- in other words: a strategy using
positive-operator-valued measurements on states of dimension $d$
can always be written as a strategy using projective measurement
on states of suitable dimension $d'>d$. Let's suppose at first
that Alice and Bob share a pure state $\ket{\Psi}_{AB}$. Upon
receiving input $x$, Alice performs the measurement defined by the
projectors $\{P\x_{a}\}$; similarly, upon receiving input $y$, Bob
performs the measurement defined by the projectors $\{P\y_{b}\}$.
We suppose that this is a winning quantum strategy. Now: if $x=0$
and Alice obtains the outcome $a\0=a$, she prepares on Bob's side
the state $\ket{\varphi_{a\0}}_B\simeq P\0_a\otimes
\I\ket{\Psi}_{AB}$; if $x=1$ and Alice obtains the outcome
$a\1=a'$, she prepares on Bob's side the state
$\ket{\varphi_{a\1}}_B\simeq P\1_{a'}\otimes \I\ket{\Psi}_{AB}$.
Since each projective measurement is a resolution of the identity,
there must exist $(a\0,a\1)$ such that
$\braket{\varphi_{a\0}}{\varphi_{a\1}}\neq 0$. In turn, given
these $(a\0,a\1)$, for each $y$ there must be at least an element
$P\y_{b\y}$ in each of Bob's measurements such that both
$\sandwich{\varphi_{a\0}}{P\y_{b\y}}{\varphi_{a\0}}$ and
$\sandwich{\varphi_{a\1}}{P\y_{b\y}}{\varphi_{a\1}}$ are not zero.
Therefore a subset of the outcomes of the quantum strategy defines
the deterministic strategy $\{a\0,a\1; b\y\}$. But if the quantum
strategy is winning, any subset of outcomes must be winning as
well: therefore, the assumption of a winning quantum strategy
implies the existence of a winning classical strategy as well,
contradicting the definition of pseudo-telepathy game.

Finally, the assumption, that Alice and Bob share a pure state,
can be easily dispensed with: if a strategy based on a mixed state
is winning, then it is winning on all the pure states onto which
the mixture can be decomposed; thus we are brought back to the
previous case, and the proof is now complete.

\end{proof}

We want to stress that this Theorem does not exclude either (i)
the existence of a pseudo-telepathy game with input cardinality
$2\times (n_{B_1}\times n_{B_2})$; nor (ii) the possibility that,
in a game that cannot be won with probability 1 (and is therefore
not a pseudo-telepathy game), the success probability may be
larger using quantum than using classical strategies. For (i): if
``Bob'' is actually Bob$_1$ and Bob$_2$, the outcomes must also of
the form $b\y=b_1\y+b_2\y$: it is this additional requirement is
what makes the pseudo-telepathy possible, as in the GHZ-Mermin
game \cite{ghz89,mermin90a}. If this requirement would be removed
(that is, if Bob$_1$ and Bob$_2$ would be allowed to communicate),
the quantum strategy still exists of course, but or theorem proves
that a classical strategy would become possible as well (again, a
well-known statement for the example of the GHZ-Mermin game). For
(ii), the CHSH Bell inequality provides such an example (see
\cite{gisin} and references therein): classical strategies succeed
with $p=\frac{3}{4}$, quantum strategies with probability
$p=\frac{2+\sqrt{2}}{4}$, but the game can be won with probability
one only allowing the ``PR-box'' \cite{pr94} as a resource.

Since games with input cardinality of $3\times 3$~\cite{aravind02}
and $2\times 2\times 2$~\cite{mermin90a} are know to exist,
Theorem \ref{thm:2xn} entails immediately the following

\begin{corol}\label{corol:main}
The minimal cardinality of the set of possible questions in
pseudo-telepathy is $3\times 3$ in the bipartite scenario and
$2\times 2\times 2$ in the multipartite.
\end{corol}

To extend our result to multipartite pseudo-telepathy, one could
reformulate the statement of Theorem~\ref{thm:2xn} as:
\begin{quotation}
For a pseudo-telepathy game to exists, every player must perceive
the number of possible combination of questions asked to the other
player(s) to be greater than two.
\end{quotation}


\section{A Bell-type inequality from the Magic Square game}\label{sec:msg}

\subsection{Definition of the game}

We present here the pseudo-telepathy game generally known as the
Magic Square game~\cite{aravind02}. The participants, namely Alice
and Bob, are each presented with a question: a random trit
$x\in\{0,1,2\}$ and $y\in\{0,1,2\}$ respectively. They must
produce three bits each, $(a_0\x,a_1\x,a_2\x)$ and
$(b_0\y,b_1\y,b_2\y)$ respectively. In order for them to win,
three requirements must be fulfilled: \ba {\cal R}_{L,A}&:&
a_0\x\oplus a_1\x\oplus a_2\x\,=\,0\,;\label{rla}\\
{\cal R}_{L,B}&:& b_0\y\oplus b_1\y\oplus b_2\y\,=\,1\,;\label{rlb}\\
{\cal R}_{AB}&:& a_{y}\x\,=\,b_{x}\y\,.\label{rab}\ea Note that
the first two requirements are ``local'', only the third one
introduces a correlation between Alice and Bob.

Without resorting to quantum mechanics or communicating, Alice and
Bob cannot win this game with probability one. In fact, for
requirement (\ref{rab}) to be always fulfilled, Alice and Bob must
draw their outcomes from a common $3\times 3$ table of 0s and 1s
with entries $c_{xy}$. But the other two requirements (\ref{rla})
and (\ref{rlb}) say that the sum of the elements in each row must
be even and the sum of the elements in each column must be odd: a
simple parity argument shows that this is impossible. For
instance, suppose Alice and Bob share a table of the form of Table
\ref{tablex}. Then they succeed for eight possible inputs, but
fail for the case $a=b=2$: in fact, requirement ${\cal R}_{L,A}$
forces $a_2^{(2)}=c_{00}\oplus c_{10}\oplus c_{01}\oplus c_{11}$,
requirement ${\cal R}_{L,B}$ forces $b_2^{(2)}=a_2^{(2)}+1$, and
requirement ${\cal R}_{AB}$ forces $a_2^{(2)}=b_2^{(2)}$.

\begin{table}
\begin{tabular}{|c|ccc|}
  \hline
  $x\setminus y$ & 0 & 1 & 2 \\
  \hline
  0 & $c_{00}$ & $c_{01}$ & $c_{02}=c_{00}\oplus c_{01}$ \\
  1 & $c_{10}$ & $c_{11}$ & $c_{12}=c_{10}\oplus c_{11}$ \\
  2 & $c_{20}=c_{00}\oplus c_{10}\oplus 1$ & $c_{21}=c_{01}\oplus c_{11}\oplus 1$ & $c_{22}=\,?$ \\
  \hline
\end{tabular}
\caption{Example of a classical table for the Magic Square game.
In the table, one reads the outputs $c_{xy}\in\{0,1\}$. Upon
receiving input $x$, Alice outputs the three bits in the
corresponding row; upon receiving input $y$, Bob outputs the three
bits in the corresponding column. The value of $c_{22}$ cannot be
chosen such as to fulfill all three requirements
(\ref{rla})-(\ref{rab}).}\label{tablex}
\end{table}

On the other hand, if Alice and Bob are allowed to share
entanglement, they exists an explicit successful strategy
\cite{aravind02}. Success takes advantage of ``contextuality'',
i.e. the fact that ``unmeasured quantities are not determined'' in
quantum physics: Alice and Bob perform only the measurements they
are asked to perform, and each pair of measurements produces
outcomes with the required relation, although the results cannot
derive from an underlying classical table of pre-determined
outcomes.

\subsection{Study of the derived Bell-type inequality}

\subsubsection{The inequality}

The classical bound for the Magic Square game can be re-written in
the form of a Bell-type inequality. Any classical strategy is the
convex mixture of extremal points which are deterministic
strategies, that is, shared tables $c_{xy}$. From the example
above in Table \ref{tablex}, it is clear that the game cannot be
won for all inputs, but can be won for eight inputs out of nine;
therefore the inequality \ba \sum_{x,y=0,1,2} \Pr\left[{\cal
R}_{L,A},{\cal R}_{L,B},{\cal R}_{AB} \vert x,y\right]&\leq& 8\,,
\label{eq:ineqmsg}\ea holds for classical strategies. In this
language, the existence of a winning quantum strategy means that
quantum physics violates this inequality up to the maximal
algebraic value of 9. This violation can be obtained the tensor
product of two maximally entangled states of two qubits
\cite{aravind02}, which is equivalent to the maximally entangled
state of two 4-dimensional systems; therefore, this inequality
does not exhibit the ``anomaly'' reviewed in \cite{ms06}. We set
out to study whether this inequality is tight, that is, a facet of
a polytope. The answer to this question requires a careful
definition of the probability space.

Let's first count how many deterministic strategies saturate the
inequality. First, Alice and Bob agree on the pair of inputs
$(x_?,y_?)$ for which they fail the game (the position of the
``?'' sign); there are 9 possibilities. Once this position fixed,
4 bits can be freely chosen and the others are determined by the
local requirements. Note that Alice's and Bob's outputs at the
position of ``?'' are fixed: in fact, for $x=x_?$ Alice must
output the bit that fulfills ${\cal R}_{L,A}$; otherwise, she will
fail as soon as $x=x_?$, independently of Bob's input, and such a
strategy does not saturate the inequality. A similar argument
holds for Bob. In conclusion, the number of deterministic
strategies that saturate the inequality is $9\times 2^4=144$.

In full generality, we are studying a class of problems in which
both Alice and Bob have 3 inputs and 8 outcomes (all the possible
three-bit lists). In this general sense, the local polytope is
therefore $\Omega_{38}$. This polytope lives in a probability
space of dimension 483; therefore, a hypersurface containing only
144 extremal points has too small dimensionality to be a facet.
However, this general view puts all the requirements of the Magic
Square game on the same footing: in particular, it supposes that
Alice and Bob may also fail to fulfill their own local
requirements ${\cal R}_{L,A}$ and ${\cal R}_{L,B}$, which is a
rather uninteresting and inefficient way of losing the game, since
it would guarantee a failure probability of at least $1/3$. It is
more natural to assume that {\em both Alice and Bob promise to
fulfill their own local requirement}, and that the only
requirement that can sometimes fail is ${\cal R}_{AB}$, which
involves correlations. In this restricted sense, for each input,
Alice and Bob have only 4 possible outcomes each, since as soon as
they choose two bits, the value of the third one is fixed. The
local polytope becomes then $\Omega_{34}$, that lives in a
probability space of dimension 99. The inequality reads now \ba
\sum_{x,y=0,1,2} \Pr\left[a_y\x = b_x\y \vert x,y,{\cal
R}_{L,A},{\cal R}_{L,B}\right]&\leq& 8\,. \label{eq:ineqmsg2}\ea
Now we can state the following

\begin{prop}
Inequality (\ref{eq:ineqmsg2}) is tight, i.e. it defines a facet
of $\Omega_{34}$.
\end{prop}

\begin{proof}
We must prove that the 144 extremal points that saturate
inequality (\ref{eq:ineqmsg2}) define a hypersurface of dimension
98. Each of the points is written as a 99-component vector,
following e.g. the parametrization introduced in \cite{cg03}
(where the vectors are written as square tables, but this ordering
is just for convenience). The 144 vectors are then arranged in a
matrix; the Proposition follows if this matrix has full rank,
because 99 linearly independent vectors are needed to define a
hyperplane of dimension 98. This verification was made with a
computer by two of us independently
and without any approximation.
\end{proof}
A remark: it is known that any facet of a polytope can always be
``lifted'' in a natural way and give a facet of a polytope with
more parties and/or more inputs and/or more outcomes
\cite{pironio}. This is not a contradiction with the fact that the
Magic Square game defines a facet of $\Omega_{34}$ and not of
$\Omega_{38}$: the lifted version of inequality
(\ref{eq:ineqmsg2}) to eight outcomes on each side is {\em not}
inequality (\ref{eq:ineqmsg}).

\subsubsection{Abstract version of the inequality}

Inequality (\ref{eq:ineqmsg2}) defines a facet of $\Omega_{34}$
which had not been listed before to our knowledge; remember that
quantum physics violates it up to the algebraic maximum. It is
instructive to rewrite the inequality by removing any reference to
the Magic Square game: after all, this inequality is a facet of
$\Omega_{34}$, it must then be possible to write it as a formula
involving only the ternary inputs and quaternary outputs.

Alice receives input $x\in\{0,1,2\}$ and produces the output
$a\x\in\{0,1,2,3\}$. We can decide conventionally that the bits
$a_0\x$ and $a_1\x$ of the Magic Square game are defined through
$a\x=2a_0\x+a_1\x$, and then $a_2\x=a_0\x\oplus a_1\x$ (but any
other convention would be valid, and would simply define an
equivalent facet). Therefore $a_0\x=\left(\frac{a\x-a\x\mod 2}{2}
\right)\mod 2$, $a_1\x=a\x\mod 2$ and
$a_2\x=\left(\frac{a\x+a\x\mod 2}{2} \right)\mod 2$. We make a
similar definition for Bob; note that
$b_2\y=\left(\frac{b\y+b\y\mod 2}{2}+1 \right)\mod 2$. Thus,
denoting equality modulo 2 by the symbol $\stackrel{2}{\equiv}$,
the inequality reads as follows: \ba \Pr\left[\frac{a\0-a\0\mod
2}{2}\,\stackrel{2}{\equiv}\, \frac{b\0-b\0\mod 2}{2}\right]+
\Pr\left[\frac{a\1-a\1\mod 2}{2}\,\stackrel{2}{\equiv}\, b\0\mod
2\right] \nonumber\\ + \Pr\left[\frac{a\2-a\2\mod
2}{2}\,\stackrel{2}{\equiv}\, \frac{b\0+b\0\mod 2}{2}+1\right] +
\Pr\left[a\0\mod 2\,\stackrel{2}{\equiv}\, \frac{b\1-b\1\mod
2}{2}\right]\nonumber\\ + \Pr\left[a\1\mod
2\,\stackrel{2}{\equiv}\, b\1\mod 2\right] +\Pr\left[a\2\mod
2\,\stackrel{2}{\equiv}\,
\frac{b\1+b\1\mod 2}{2}+1\right]\nonumber\\
\Pr\left[\frac{a\0+a\0\mod 2}{2}\,\stackrel{2}{\equiv}\,
\frac{b\2-b\2\mod 2}{2}\right]+ \Pr\left[\frac{a\1+a\1\mod
2}{2}\,\stackrel{2}{\equiv}\, b\2\mod 2\right]  \nonumber\\ +
\Pr\left[\frac{a\2+a\2\mod 2}{2}\,\stackrel{2}{\equiv}\,
\frac{b\2+b\2\mod 2}{2}+1\right]\,\leq \, 8\,. \label{ineqimpl}\ea
This is how the inequality looks like (up to symmetries) if the
outputs of Alice and Bob appear explicitly as quaternary values:
the expression is less elegant than Eq.~(\ref{eq:ineqmsg2}), and
the origin is completely hidden.

\subsection{Error tolerance}

For $n_A=n_B=4$, unstructured noise is the probability distribution
$P(a,b|x,y)=\frac{1}{16}$ (random uncorrelated outputs for any
input). The amount $p_n$ of unstructured noise that one needs to add
to the quantum correlations in order to make them local is a measure
of non-locality called ``resistance to noise''
~\cite{werner89,ms06}. Denoting $I_{QM}$, $I_{LV}$ and
$I_{\text{noise}}$ the values that an inequality achieves with QM,
local variables and unstructured noise respectively, $p_n$ is
defined by the linear relation
$(1-p_n)I_{QM}+p_nI_{\text{noise}}=I_{LV}$ that is \ba
p_n&=&\frac{I_{QM}-I_{LV}}{I_{QM}-I_{\text{noise}}}\,. \ea For our
new inequality (\ref{eq:ineqmsg2})-(\ref{ineqimpl}), we know
$I_{QM}=9$ and $I_{LV}=8$, and it is easy to compute that
$I_{\text{noise}}=\frac{9}{2}$: in fact, since bits computed from
random $a$ and $b$ are random bits, all the nine terms of the
inequality become just the probability that two random bits are
equal. All in all, we obtain $p_n = \frac{2}{9}=0.\bar{2}$.

This value can be compared to the value obtained for other
inequalities with the same number of outcomes. The best known one is
$I_{2244}$ \cite{cg03}. For this inequality, one has $I_{LV}=0$,
$I_{\text{noise}}=-\frac{3}{4}$ and $I_{QM}\approx 0.3648$
--- this last value comes from the value $I'_{QM}=2.9727$ given in Table I of
\cite{adgl02} for the equivalent inequality $I'$ defined by
$I=\frac{4-1}{2\times 4}(I'-2)$. Therefore $p_n\approx 0.3272$. We
see that an inequality, which is not related to any pseudo-telepathy
game (because of Theorem \ref{thm:2xn} in this paper, has a stronger
resistance to noise than the one based on the Magic Square game.
Another inequality that can be checked is $I_{3344}$; we have done
it numerically and found a smaller $p_n$ than for our new
inequality.

In summary, no clear-cut picture can be derived from the
comparison of inequalities based on the resistance to noise (see
also the discussion in \cite{adgl02}); but to our present
knowledge, pseudo-telepathy does not provide the best possible
resistance to noise.


\section{Conclusion}~\label{sec:conc}

With the first part of this study (Section \ref{sec:min}), we now
have a complete picture of the minimal requirements for
pseudo-telepathy:
\begin{itemize}
\item Bipartite scenario.
\begin{itemize}
\item Minimal entanglement: $3\times 3$~\cite{bmt04}. \item
Minimal output cardinality: $3\times 2$~\cite{chtw04}. \item
Minimal input cardinality: $3\times 3$.
\end{itemize}
\item Multipartite scenario.
\begin{itemize}
\item Minimal entanglement: $2\times 2\times 2$. \item Minimal
output cardinality: $2\times 2\times 2$. \item Minimal input
cardinality: $2\times 2\times 2$.
\end{itemize}
\end{itemize}
It is somewhat peculiar that there is so much symmetry in the
minimal requirements, up to the 2 in the minimal output cardinality.
It would thus be interesting to know why we can ``save'' only in
output cardinality. To further understand pseudo-telepathy, we would
need to characterize the potential role of POVMs in this framework.
While they have been taken into account for the studies of the
minimal requirements discussed here, it is not clear whether the use
of POVMs could lower the maximal probability of a classical strategy
to win an instance of a pseudo-telepathy game for certain fixed
dimensions. Another open question of interest is whether we can find
a bipartite pseudo-telepathy game which only uses the minimal
requirements or if there some kind of trade-off between these
properties.

The second part of this work is a contribution to the study of
pseudo-telepathy games in the broader context of non-local
probability distributions. We have shown that a tight inequality
is associated to the bipartite Magic Square game, but the question
remains open, whether all pseudo-telepathy games define tight
Bell-type inequalities. The resistance to noise of this new
inequality is not the highest known to date for four-dimensional
outputs.


\section*{Acknowledgements}
We are grateful to Cyril Branciard, Nicolas Brunner and Alain Tapp
for stimulating discussions. This work has been supported by the
European Commission under the Integrated Project Qubit
Applications (QAP) funded by the IST directorate as Contract
Number 015848 and the Swiss NCCR Quantum Photonics.


\begin{thebibliography}{99}

\bibitem{hr83}
P.~{Heywood} et M.~L.~G.~{Redhead}, ``Nonlocality and the
{K}ochen-{S}pecker paradox'', \textit{Foundations of Physics}
\textbf{13}: 481--499, 1983.



\bibitem{ghz89} D.~M.~{Greenberger}, M.~A.~{Horne} and
A.~{Zeilinger}, ``Going beyond {B}ell's theorem'', in {\it Bell's
Theorem, Quantum Theory, and Conceptions of the Universe}, edited
by M. Kafatos (Kluwer Academic, Dordrecht), pages~69--72, 1989.



\bibitem{mermin90a}
N.~D.~{Mermin}, ``Quantum mysteries revisited'', \textit{American
Journal of Physics} \textbf{58}: 731--743, 1990.


\bibitem{bct99}
G.~{Brassard}, R.~{Cleve} and A.~{Tapp}, ``Cost of exactly
simulating quantum entanglement with classical communication'',
\textit{Physical Review Letters} \textbf{83}: 1874--1877, 1999.






\bibitem{bell64}
J.\,S.~Bell,
\newblock ``On the {E}instein-{P}odolsky-{R}osen paradox'',
\newblock \textit{Physics} \textbf{1}: 195--200, 1964.



\bibitem{hardy92}
L.~{Hardy}, ``Quantum mechanics, local realistic theories, and
{L}orentz-invariant realistic theories'', \textit{Physical Review
Letters} \textbf{68}: 2981--2984, 1992.



\bibitem{bmt04}
G.~{Brassard}, A.~A.~{M\'ethot} and A.~{Tapp}, ``Minimal bipartite
state dimension required for pseudo-telepathy'', \textit{Quantum
Information and Computation} \textbf{5}: 275--284, 2005.



\bibitem{methot05}
A.~A.~{M{\'e}thot}, ``On local-hidden-variable no-go theorems'',
\textit{Canadian Journal of Physics}, to appear, preprint
available at http://arxiv.org/quant-ph/0507149.



\bibitem{bbt04a}
G.~{Brassard}, A.~{Broadbent} and A.~{Tapp}, ``Quantum
pseudo-telepathy'', \textit{Foundations of Physics} \textbf{35}:
1877--1907, 2005.



\bibitem{chtw04}
R.~{Cleve}, P.~{H\o yer}, B.~{Toner} and J.~{Watrous},
``Consequences and limits of nonlocal strategies'',
\textit{Proceedings of 19th IEEE Conference on Computational
Complexity}: 236--249, 2004.

\bibitem{pito} I.~Pitowski, {\em Quantum Probability, Quantum Logic}, Lecture Notes in Physics {\bf 321} (Springer Verlag,
Heidelberg, 1989)

\bibitem{cg03}
D.~Collins, N.~Gisin, ``A relevant two qubit {B}ell inequality
inequivalent to the {CHSH} inequality'', \textit{Journal of
Physics A: Mathematical and General} \textbf{37}: 1775--1787,
2004.

\bibitem{proofmermin} R. F. Werner and M. M. Wolf, ``All-multipartite
Bell-correlation inequalities for two dichotomic observables per
site'', \textit{Physical Review A} \textbf{64}: 032112, 2001.

\bibitem{aravind02}
P.~K.~{Aravind}, ``Bell's theorem without inequalities and only
two distant observers'', \textit{Foundations of Physics Letters}
\textbf{15}: 397--405, 2001.


\bibitem{gisin} N.~Gisin, ``Can relativity be considered complete?
>From Newtonian nonlocality to quantum nonlocality and beyond'',
preprint available at http://arxiv.org/quant-ph/0512168.

\bibitem{pr94}
S.~Popescu and D.~Rohrlich, ``Quantum Nonlocality as an Axiom'',
\textit{Found. Phys.} \textbf{24}: 379--385 (1994).

\bibitem{ms06}
A.~A.~{M\'ethot} and V.~Scarani, ``An anomaly of non-locality'',
preprint available at http://arxiv.org/quant-ph/0601210.





\bibitem{pironio}
S.~Pironio, ``Lifting Bell Inequalities'', \textit{Journal of
Mathematical Physics} \textbf{46}: 062112, 2005.


\bibitem{werner89}
R.~F.~{Werner},  ``Quantum states with
{E}instein-{P}odolsky-{R}osen correlations admitting a
hidden-variable model'', \textit{Physical Review A} \textbf{40}:
4277--4281, 1989.


\bibitem{adgl02}
A.~Ac\'{\i}n, T.~Durt, N.~Gisin and J.~I.~Latorre, ``Quantum
nonlocality in two three-level systems'', \textit{Physical Review
A} \textbf{65}: 052325, 2002.




\end{thebibliography}
\end{document}